\documentclass[11pt,a4paper]{article}
\usepackage{jheppub}

\usepackage{hyperref}

\usepackage{epsfig}
\usepackage{amsmath,latexsym,amssymb}
\usepackage{graphicx}
\usepackage[latin1]{inputenc}
\usepackage{braket}
\usepackage{pdfsync}
\usepackage{definitions}

\usepackage{framed}

\usepackage{mathtools}

\def\be{\begin{equation}}
\def\ee{\end{equation}}
\def\ba{\begin{eqnarray}}
\def\ea{\end{eqnarray}}

\newcommand{\lp}{\left(}
\newcommand{\rp}{\right)}

\newcommand{\apr}{\alpha'}

\newcommand{\bPhi}{\bar{\Phi}}
\newcommand{\bz}{\bar{z}}

\newcommand{\bX}{\widetilde{X}}

\newcommand{\bpsi}{\bar{\psi}}

\newcommand{\pS}{ \partial \Sigma }

\def\Mo{{\cal M}_{0,N}}
\def\sv{{\rm sv}}

\def\SV{{\zeta_{\rm sv}}}

\def\Hc{{\cal H}}

\def\IQ{{\bf Q}}
\def\IN{{\bf N}}
\def\IR{{\bf R}}
\def\IC{{\bf C}}
\def\IP{{\bf P}}

\def\sv{{\rm sv}}

\def\SV{{\zeta_{\rm sv}}}

\def\IP{{\bf P}}\def\IC{{\bf C}}\def\IR{ {\bf R}}
\def\IN{ {\bf N}}
\def\IQ{ {\bf Q}}

\def\z{\zeta}

\def\cP{{\cal P}}

\newcommand{\comment}[1]{}

\def\fc#1#2{{\frac{#1}{#2}}}
\newcommand{\req}[1]{(\ref{#1})}

\def\Hc{{\cal H}}

\def\ap{{\alpha'}}

\newcommand{\eea}{\end{eqnarray}}

\def\z{\zeta}

\def\ds{\displaystyle}

\def\sv{{\rm sv}}

\title{\boldmath SV--MAP BETWEEN TYPE I AND HETEROTIC SIGMA MODELS\unboldmath}

\abstract{The scattering amplitudes of gauge bosons in heterotic and open superstring theories are related by the single--valued projection which yields heterotic amplitudes by selecting a subset of multiple zeta value coefficients in the $\alpha'$ (string tension parameter) expansion of  open string amplitudes. In the present work, we argue that this relation holds also at the level of low--energy expansions (or individual Feynman diagrams) of the respective effective actions, by investigating the beta functions of two--dimensional sigma models describing world--sheets of open and heterotic strings. We analyze the sigma model Feynman diagrams generating identical effective action terms in both theories and show that the heterotic coefficients are given by the single-valued projection of the open ones. The single-valued projection appears as a result of summing over all radial orderings of heterotic vertices on the complex plane representing string world-sheet.}

\preprint{MPP--2017--194}
\keywords{superstring theory, sigma models, scattering amplitudes, multiple zeta values}

\author{Wei Fan${}^{1,4}$ ,
A. Fotopoulos${}^{1,3}$, S. Stieberger${}^{2}$,
T.R. Taylor${}^{1,4}$ \\[0.5cm]

   $^1$ {\it Department of Physics \\
  Northeastern University, Boston, MA 02115, USA}\\[0.2cm]

 $^2$  {\it Max--Planck--Institut f{\"u}r Physik\\
 Werner-Heisenberg-Institut, 80805 M{\"u}nchen, Germany}\\[0.2cm]

 $^3$ {\it Department of Applied Sciences \\
  Mount Ida College, Newton 02159, USA}\\[0.2cm]

 $^4${ \it Institute of Theoretical Physics, Faculty of Physics \\
University of Warsaw, Poland}

\bigskip
}

\begin{document}
\maketitle

\section{Introduction} \label{intro}

Perturbative open and closed string amplitudes seem to be rather different due to distinct underlying world--sheet topologies, with or without boundaries. Their explicit computations however, reveal some unexpected connections, in particular
the Kawai--Lewellen--Tye (KLT) relations \cite{KLT}.
At the  tree level, the amplitudes describing the scattering of open string states appear from a disk world--sheet, with the
vertex operator insertions giving rise to  real iterated integrals at the boundary.
In closed string theory, the vertices are inserted and integrated on the
complex sphere. The latter integrals can be expressed in terms of a sum over squares of open string amplitudes
by using the KLT method \cite{KLT}.
In~\cite{Stieberger:2014hba}, another relation  between open and closed  string tree--level amplitudes has been found. Generally, complex world--sheet integrals can be expressed as real iterated integrals by means of  the  single--valued projection (\sv) to be explained below.
As a consequence, and in contrast to \cite{KLT}, tree--level closed string amplitudes can be expressed in terms  of  single--valued projections of open string amplitudes.
Recently, it has been argued that similar relations are expected when comparing open and closed string one--loop amplitudes \cite{Brown:2017qwo}.

In \cite{Stieberger:2014hba} two of the present authors demonstrated that the single trace part of the $N$--point tree--level heterotic superstring amplitudes ${\cal A }^{HET}_N$ is given by the single--valued  projection of the corresponding type--I amplitude ${\cal A}^I_N$:
\be\label{sv}
{\cal A }^{HET}_N={\rm sv}({\cal A}^I_N)\ .
\ee
The amplitudes  ${\cal A}_N$  can be expanded in the inverse string tension $\ap$, yielding rational functions of the kinematic invariants multiplied by periods on $\Mo$ giving rise to multiple zeta values (MZVs), see Ref. \cite{Stieberger:2016xhs} for more details and references therein\footnote{$\Mo$ describes the moduli space of  Riemann spheres (genus zero curves) with
$N\geq 4$ ordered marked points  modulo the action $PSL(2,\IC)$ on those points.}.
From \req{sv} it follows that the $\ap$--expansion of the closed string amplitude can be obtained from that of the open superstring amplitude by simply replacing MZVs by their corresponding single--valued
multiple zeta values (SVMZVs) according to the rules of single-valued projection~\sv.

The $\ap$--expansion of the scattering amplitudes is related to two--dimensional sigma models, in the following way.
The scattering of massless gauge bosons can be described by an effective action that contains, in addition to the Yang--Mills term, an infinite series of interactions appearing order by order in $\ap$. It is a low energy expansion that includes the effects of heavy string modes. This action generates effective field equations. On the other hand,
dynamics of strings propagating in gauge field backgrounds are described by two--dimensional sigma models.
 The world--sheet conformal invariance of strings propagating in  gauge field backgrounds requires, among other things,  the vanishing of the beta function associated to the coupling of background fields to the string world--sheet. This requirement leads to background field equations that should be equivalent to the equations generated by the effective action.  The corresponding beta functions can be computed in sigma model perturbation theory, by using Feynman diagrams.
In this context, $\ap$ is the sigma model coupling, with $\ap^{\ell}$ appearing at the $\ell$th--loop order.

In this work, we show that the $\sv$--projection can be applied to {\it individual\/} sigma model Feynman diagrams with appropriately regulated ultra--violet divergences\footnote{From the physical point of view, SVMZVs first have appeared in the computation of graphical functions (positive functions on the punctured complex plane) for certain Feynman diagrams \cite{Schnetz} and then rigorously defined in \cite{SVMZV}.}. We  conjecture  that the single--valued projection of the  beta function of type--I theory gives the beta--function of the heterotic string. We support this conjecture by explicit three and four--loop computations  in open and heterotic sigma models. As a corollary, the effective low--energy action describing gauge fields in heterotic superstring theory can be obtained from the respective open string action by replacing MZVs by SVMZVs.

The paper is organized as follows.
In Section \ref{open}, we setup the conventions and recall some basic results of  previous computations in type I open superstring theory, in particular the one and two--loop beta functions. In Section \ref{heterotic}, we setup the heterotic string perturbation theory. We compute the one--loop beta function by using standard world-sheet Feynman diagrams. Then we demonstrate how we can reorganize perturbation theory in terms of a background gauge covariant action which is in complete parallel with the type-I perturbation theory and use it to reproduce the known result that the two--loop beta function is vanishing.
In Section \ref{svmapprop}, we describe some features of single--valued multiple zeta values and propose a general sv--map for the heterotic string.
It connects the beta function of the type--I superstring with the heterotic one via a simple application of the sv--map:
$\beta_h  = {\rm sv}(\beta_o)$.
In Section \ref{het3loop}, we discuss three--loop Feynman diagrams contributing to the beta functions.
 We introduce the integration variables and a regularization scheme such that the \sv--map applies at the level of {\it individual} Feynman diagrams. In Section \ref{het4loops} we compute three representative four--loop Feynman diagrams, in each case finding that the heterotic integrals  are given by the \sv\ projection of real, ordered open string integrals. Our conclusions are summarized in Section \ref{conclusions}. In the Appendix, we outline a method for performing the angular part of Feynman loop integrals encountered in the heterotic sigma model.

\section{Open superstring sigma model and two--loop beta function}\label{open}
The purpose of this section is to establish notation and to recall some basic results of sigma-model computations in open superstring theory.

The action describing the world-sheet $X^\mu(\sigma_1,\sigma_2)$ of open strings propagating in a general non-abelian gauge background $A_\mu(X)$ contains the bulk and boundary contributions \cite{DO, BP}:
\be \label{openaction}
S=S_{\Sigma} + S_{\partial \Sigma}\ee
with the bulk action
\be S_\Sigma= {1\over 4 \pi \apr}  \int d^2\sigma  [\sqrt{\gamma} \gamma^{a b} \partial_a X^\mu \partial_b X_\mu -{i \over 2} \bPhi^\mu \rho^a \partial_a \Phi_\mu ] \ee
and the boundary action
\be
S_{\partial \Sigma}= \ln \ Tr \ \cP ( U[A])\ee
determined by the Wilson loop
\begin{equation}
U[A]=\ exp \big(i g \oint_{\pS} d \tau \big[A_\mu(X) \partial_\tau X^\mu- {1\over 2} \phi^\mu \phi^\nu F_{\mu \nu} \big] \big)\ .
\end{equation}
In this theory, $\Phi^\mu$ are the fermionic coordinates, with $\phi^\mu=\Phi^\mu|_{\partial\Sigma}$. We are using the notation of Ref.~\cite{BP}.

The main focus of this study are the ultraviolet singularities due to quantum fluctuations of string coordinates and their fermionic partners around the classical background. The bosonic coordinates are expanded around the classical background  $\widetilde{X}$ as $X= \widetilde{X} +\xi$. The respective expansion of the Wilson loop is \cite{Gervais:1979zp} (see also \cite{BP}):
\ba
&&~~~~~~~~~~~~~S_{\partial \Sigma}(\widetilde{X} +\xi) ~=~  ig \oint d\tau \, Tr \ \cP \bigg( U [ A] \bigg[ A_\mu(\widetilde X) \partial_\tau \widetilde X^\mu \nonumber\\[2mm]&&  + \, F_{\mu\rho} \xi^\mu \partial_\tau \widetilde X^\rho -{1\over 2} F_{\mu\nu} \phi^\mu \phi ^\nu
- {1\over 2} D_\rho F_{\mu\nu} \xi^\rho \phi^\mu \phi ^\nu +{1\over 2} D_\mu F_{\nu \rho}\xi^\mu \xi^\nu \partial_\tau \widetilde X^\rho  \label{Wilsonexpansion}  \\
&&+\, {1 \over 2} F_{\mu\nu} \xi^\mu \partial_\tau \xi^\nu -{1\over 4} D_\rho D_\sigma  F_{\mu\nu} \xi^\rho \xi^\sigma \phi^\mu \phi ^\nu+  \sum_{n=3}^\infty \bigg( {1\over n!}D_{\mu_1} \dots D_{\mu_{n-1}} F_{\mu_n\rho}\xi^{\mu_1} \dots  \xi^{\mu_{n}} \partial_\tau \widetilde X^\rho  \nonumber\\
&&+ {n-1 \over n!} D_{\mu_1} \dots D_{\mu_{n-2}} F_{\mu_{n-1}\mu_n}\xi^{\mu_1}\dots  \xi^{\mu_{n-1}} \partial_\tau \xi^{\mu_n}  -{1\over 2} {1\over n!} D_{\mu_1} \dots D_{\mu_{n}} F_{\rho\sigma}\xi^{\mu_1}\dots  \xi^{\mu_{n}} \phi^\rho \phi^\sigma \bigg) \bigg] \bigg) \nonumber
\ea
The loop expansion leads to background-dependent ultraviolet divergences originating from the boundary couplings. Their treatment, in particular the renormalization procedure, are rather complicated. We focus on quantum corrections to the boundary gauge coupling $A_\mu(X) \partial_\tau X^\mu$. It is known that the requirement of the vanishing of the associated field-dependent beta function, order by order in the string tension $\alpha'$, is equivalent to background field equations \cite{NLSM, AG1981}. In ambient space--time, the corresponding $\alpha'$--dependent corrections to Yang--Mills action are due to heavy string modes, integrated out at low energies.

In sigma model perturbation theory, the basic quantity is the free propagator of bosonic fluctuations $\xi$:
\be\label{openprop}\displaystyle
G(\tau,\sigma;\tau',\sigma')=G(z,z')= -\frac{\apr}{2} \ln |z-z'|^2 -\frac{\apr}{2} \ln |z-\bar z'| ^2\ .
\ee
In Feynman diagrams, this propagator will be represented by a wavy line. Among Feynman rules, it is also convenient to include the propagator marked by a slash, with the mark representing the derivative of $G(z,z')$ with respect $z$ or $z'$, whichever point is closer to the mark. There is an infinite set of gauge field-dependent interactions represented by Feynman vertices.
\begin{figure}[b]
\begin{center}
\includegraphics[scale=0.9]{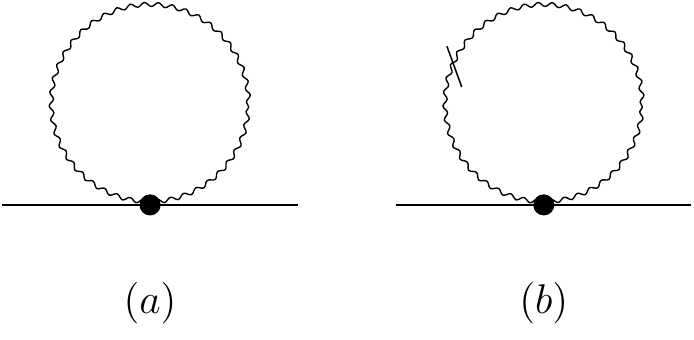}
\caption{The solid line represents the world-sheet boundary. The wiggly line represents the bosonic propagator and the slash a derivative. Diagram (b) vanishes due to the symmetry properties of the propagator.}\label{figopen1loop}
\end{center}
\end{figure}

At one--loop, the coupling under consideration receives loop corrections shown in Figure \ref{figopen1loop}. Diagram (a) contains a short-distance singularity $\lim_{\tau\to\tau'}G(\tau-\tau')$.
It can be regulated by introducing a cutoff $|\tau-\tau'|>\epsilon$, so that $G(0)\to -2\alpha'\ln\epsilon$. Diagram (b) contains $\lim_{\tau\to\tau'}\partial_\tau G(\tau-\tau')$ which vanishes if one takes a symmetric combinations of the limits $\tau-\tau'=\pm \epsilon$.
As a result, in the notation of \cite{DO, BP}:
\be\label{open1loopdiag}
\makebox{Figure  \ \ref{figopen1loop}}= ig \,\apr \ln\epsilon\int Tr \ \cP \lp U [ A] D^\mu F_{\mu \rho} \partial_\tau \bX^\rho \rp\ .
\ee
At the one--loop level, fermion loops do not contribute because of symmetric limit mentioned above.
The one--loop divergence can be removed by redefining the background field $A_\rho\to A_\rho+\delta A_\rho$, with
\be\label{open1loopcounter}
\delta  A^{(1)}_\rho= \apr g D^\mu F_{\mu \rho} \ln\epsilon+ {\cal O}(\apr^2)
\ee
which cancels the logarithmic divergence.
The beta function associated to the  boundary  coupling $A_\mu(X) \partial_\tau X^\mu$
is defined as
\be\label{open1loopbeta}
\beta_\rho={\partial \over \partial(\ln \epsilon)}\delta A_\rho= \apr g D^\mu F_{\mu \rho} + {\cal O}(\apr^2)\ .
\ee
At the leading order, zero beta function requires the background to satisfy Yang-Mills field equations.

\begin{figure}[t]
\begin{center}
\includegraphics[scale=0.8]{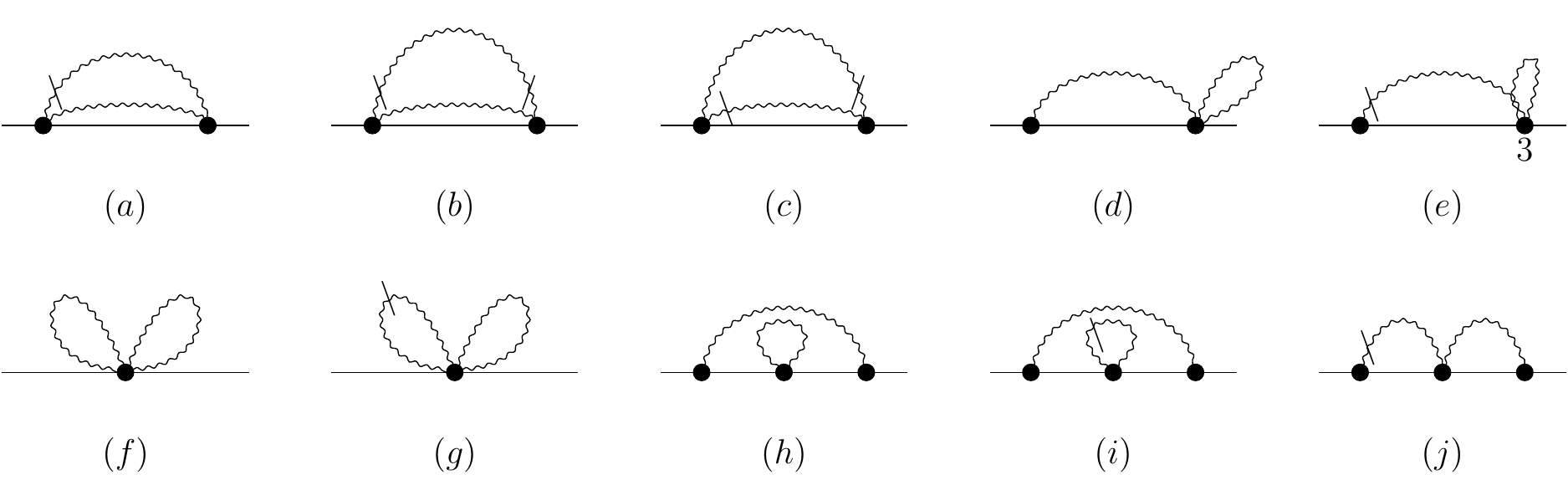}
\caption{Two--loop diagrams with bosonic loops}\label{figopen2loop}
\end{center}
\end{figure}
\begin{figure}
\begin{center}
\includegraphics[scale=0.9]{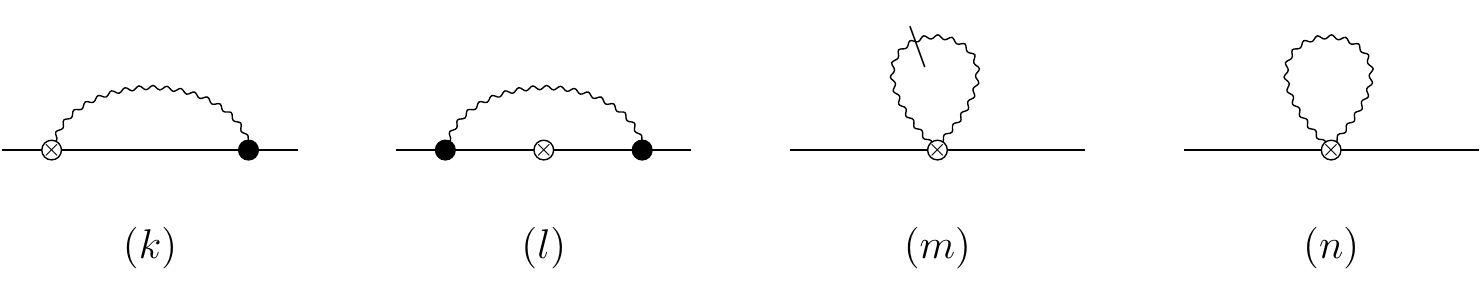}
\caption{One--loop diagrams with one--loop counterterm vertices  which contribute at two--loops}\label{figopen2loopb}
\end{center}
\end{figure}
\begin{figure}
\begin{center}
\includegraphics[scale=0.9]{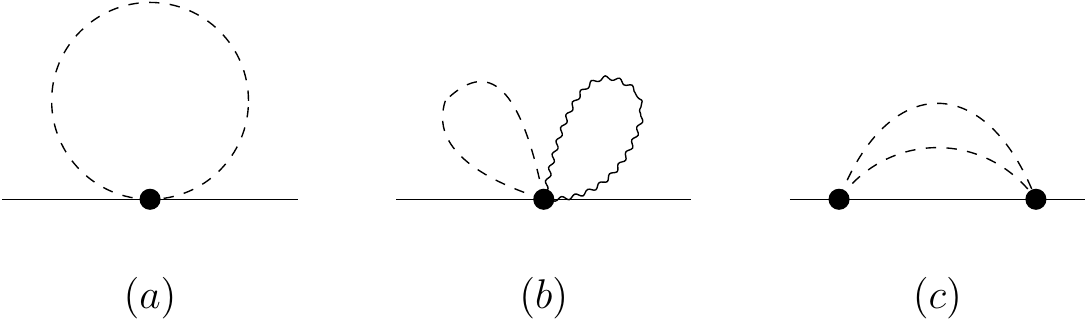}
\caption{Two--loop diagrams with fermion loops. Dashed lines are fermionic propagators. (a) and (b) are identically zero.}\label{figopen2loopferm}
\end{center}
\end{figure}
A similar analysis can be repeated at the two--loop level, with the Feynman diagrams
shown in Figure \ref{figopen2loop} and \ref{figopen2loopb}, see \cite{DO}. There are also diagrams involving fermion loops, shown in Figure \ref{figopen2loopferm}. We refer to \cite{DO} for details, however we choose to show these diagrams because  we will later make a connection to similar diagrams in the heterotic case. Note that in the open string case, the solid line is incorporated in order to remind us that the vertices are located at the boundary, while in the heterotic case, a similar line will represent a propagating fermion.

{}From the bosonic diagrams of Figures \ref{figopen2loop} and \ref{figopen2loopb}, one obtains
\be\label{beta2loopbos}
\beta^{\rm bos}_\rho={\partial \over \partial(\log \epsilon)}\delta A_\rho= \apr g D^\mu F_{\mu \rho} + i (\apr g)^2 [D_\rho F^{\mu \nu} , F_{\mu \nu}]+{\cal O}(\alpha'^3)\ .
\ee
After including fermion loops of Figure \ref{figopen2loopferm}, one finds that they cancel the bosonic contributions, leaving no ultraviolet divergences at the two--loop level:
\be\label{beta2loopopen}
\beta^{open}_\rho={\partial \over \partial(\log \epsilon)}\delta A_\rho= \apr g D^\mu F_{\mu \rho}  + {\cal O}(\apr^3)\ .
\ee
Corrections of order ${\cal O}(\apr^3)$ correspond to three--loop effects and are expected to be non-vanishing.

\section{Heterotic string sigma model and two--loop beta function}\label{heterotic}
\subsection{Heterotic sigma model}
We now proceed to one--loop and two--loop effects of the heterotic string. The heterotic string action in the presence of a background $A^{ij}_\mu(x)$ field,  is given by \cite{Sen1985}:
\begin{equation} \label{hetaction}
 \mathcal{S} = {1\over 2 \pi \apr} \int d^2z\ \Big[ 2 \partial X^{\mu }\bar{\partial }X_{\mu }
-i \phi ^{\mu } \bar{\partial }\phi _{\mu }+i\psi^j\partial \psi ^j+\psi \left(\partial X^{\nu }A_{\nu }+\frac{i}{4}F_{\mu \nu}\phi ^{\mu}\phi ^{\nu} \right)\psi \Big]
\end{equation}
where $\psi^i$  are right-handed Majorana-Weyl fermions with  gauge indices $i$ in the $SO(32)$ group or the $SO(16) \times SO(16)$ subgroup of $E_8 \times E_8$ and $\phi^\mu$ the fermionic coordinates of the superstring.  The classical field equation are
\begin{eqnarray} \label{heteqm}
 \bar{\partial} \phi_{\mu} &=&- \frac{1}{4}\psi  F_{\mu\nu} \phi ^{\nu} \psi \nonumber \\
 4 \partial \bar{\partial } X_{\mu } &=& \psi \left(F_{\mu \nu }\partial X^{\nu }+\frac{i}{4}D_{\mu }F_{\rho \sigma}\phi ^{\rho}\phi ^{\sigma}\right)\psi \\
 i\partial \psi &=& -\left(\partial X^{\nu}A_{\nu }+\frac{i}{4}F_{\mu \nu}\phi ^{\mu}\phi ^{\nu}\right)\psi\ .  \nonumber
\end{eqnarray}
As in the open string case, we expand the bosonic field around the classical background: $ X^{\mu }\to \bX^{\mu }+\xi^{\mu }$.  Fermions are treated as in ordinary perturbation theory. The action becomes
\begin{equation}
 \mathcal{S}=\mathcal{S}_0+\mathcal{S}_I
\end{equation}
where \(\mathcal{S}_0\) is the free part:
\begin{equation}
\mathcal{S}_0={1\over 2 \pi \apr} \int d^2 z \ \left(2 \partial \bX^{\mu }\bar{\partial }\bX_{\mu }+2 \partial \xi^{\mu }\bar{\partial }\xi_{\mu }-i \phi ^{\mu } \bar{\partial }\phi _{\mu }+i\psi^j\partial \psi ^j   \right)
\end{equation}
and \(\mathcal{S}_I\) is the interacting part:
\ba\label{hetSI}
 \mathcal{S}_I &=& {1\over 2 \pi \apr} \int d^2 z \ \Big[ -4 \partial \bar{\partial }\bX_{\mu }\xi^{\mu }+\psi \left(\partial \bX^{\nu }A_{\nu }+\frac{i}{4}F_{\mu \nu}\phi ^{\mu}\phi ^{\nu}\right)\psi \nonumber \\
 &+&\psi
\left(\left(\partial \bX^{\nu }A_{\nu ,\mu_1 }+\frac{i}{4}F_{\mu \nu, \mu_1}\phi ^{\mu}\phi ^{\nu}\right) \xi^{\mu_1 } +\partial \xi^{\mu }A_{\mu
}\right)\psi \nonumber \\
 &+& \frac{1}{2}\psi \left(\left(\partial \bX^{\nu }A_{\nu ,\mu _1\mu _2}+\frac{i}{4}F_{\mu \nu, \mu_1\mu_2}\phi ^{\mu}\phi ^{\nu}\right) \xi^{\mu
_1}\xi^{\mu _2}+A_{\mu _1,\mu _2}\partial \xi^{\mu _1}\xi^{\mu _2}+A_{\mu _2,\mu _1}\xi^{\mu _1}\partial \xi^{\mu _2}\right)\psi +\cdots  \nonumber\\
 &+&\ \sum_{j=1}^n \frac{1}{n!}\psi \bigg( (\partial \bX^{\nu }A_{\nu ,\mu _1\mu _2\ldots\mu _n}+\frac{i}{4}F_{\mu\nu,\mu _1\mu _2\ldots\mu _n} \phi^{\mu}\phi ^{\nu})   \xi^{\mu _1}\xi^{\mu _2}\cdots \xi^{\mu _n} \nonumber\\
&+& A_{\mu _j,\mu _1\ldots\hat{\mu}_j\ldots\mu _n}\xi^{\mu
_1}\cdots \partial \xi^{\mu _j}\cdots \xi^{\mu _n} \bigg) \psi
~+~\cdots \Big]
\end{eqnarray}
where a hat over index indicates that it is missing. Clearly, the expression given above is not invariant under background gauge transformations. In the case of closed string background of gravitational fields, one can expand the action in terms of bosonic normal coordinates to restore general coordinate invariance. In the case of the background gauge fields, this would require an equivalent ``normal" expansion of the fermionic fields $\psi^i$, but we will not follow along this route.

\begin{figure}
\begin{center}
\includegraphics[scale=1]{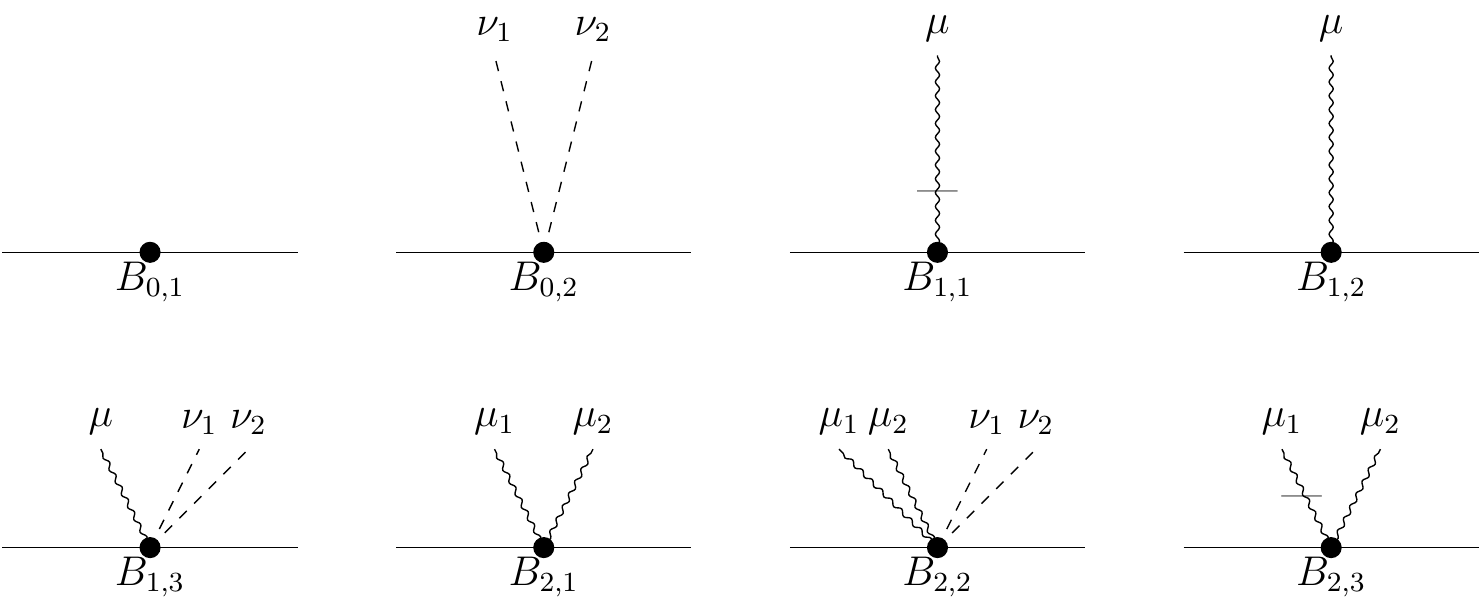}
\caption{\label{fighetvert}Solid lines correspond to $\psi$ fermions.  Dashed lines with greek letter indices correspond to the $\phi^\mu$ fermions.}
\end{center}\vskip -1 cm
\end{figure}

\begin{figure}
\begin{center}
\includegraphics[scale=1]{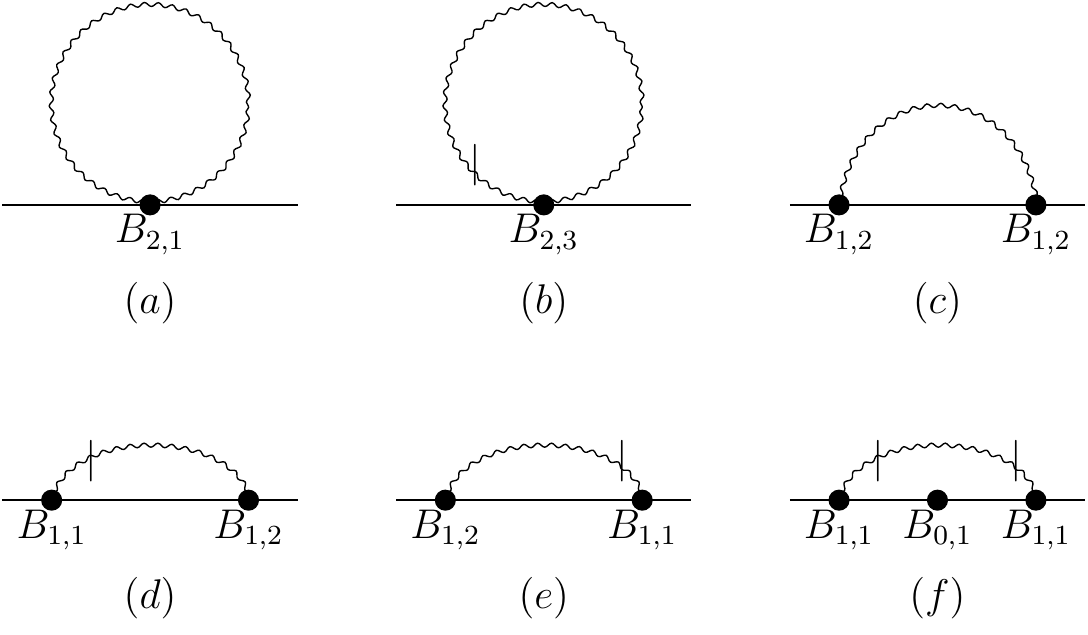}
\caption{\label{fighet1loop} One--loop contributions to the heterotic beta function}
\end{center}\vskip -1 cm
\end{figure}

 The quantum field propagators are given by:
\begin{eqnarray}\label{hetprop}
 \langle\xi^{\mu _1}\left(z_1\right) \xi^{\mu _2}\left(z_2\right)\rangle &=& \eta ^{\mu _1\mu _2}G\left(z_1,z_2\right) \\ \nonumber
 \langle\psi ^{j_1}\left(z_1\right) \psi ^{j_2}\left(z_2\right)\rangle &=& \delta ^{j_1j_2}K\left(z_1,z_2\right) \\
 \langle\phi ^{\mu _1}\left(z_1\right) \phi ^{\mu _2}\left(z_2\right)\rangle &=& \eta ^{\mu _1\mu _2}\bar{K}\left(z_1,z_2\right) \nonumber
\end{eqnarray}
where the bosonic propagator
\begin{equation}\label{hetpropmom}
 G\left(z_1,z_2\right) = -{\alpha ' \over 2}  \ln \big(\left| z_1-z_2\right|^2\big) \end{equation} is subject to \(4\partial \bar{\partial }G\left(z_1,z_2\right)=-2\pi \alpha ' \delta ^2\left(z_1-z_2\right)\). The fermionic propagators are
 \begin{equation}
 K\left(z_1,z_2\right) = 2i \bar{\partial }_1G\left(z_1,z_2\right) ~, \qquad  \bar{K}\left(z_1,z_2\right)= -2i \partial _1G\left(z_1,z_2\right)\ .
\end{equation}
As in the open string case all loop divergences can be regulated by  introducing a UV cut-off $|z-z'|>\epsilon$. Some loop diagrams require also IR cut-offs, but this will not be important for the computation of the beta function.
As mentioned above, unlike the open string case, where (\ref{Wilsonexpansion}) is background gauge invariant, the heterotic string expansion (\ref{hetSI}) is not. We will compute the one--loop correction to the heterotic string to  demonstrate how target space gauge invariance can be restored. In the background field method we compute the one--loop diagrams with external  $\psi$-fermionic fields and the corresponding counterterms. The vertices required for the one--loop computation, as well as for our discussion of gauge invariance below, can be read from (\ref{hetSI}). They have the form $\psi( B_{a,b}) \psi$ with the vertex functions $B_{a,b}$
\ba\label{hetvert}
&& B_{0,1}= A_\rho \partial \bX^\rho \qquad B_{0,2}= {i \over 4} F_{\nu_1 \nu_2}\phi^{\nu_1} \phi^{\nu_2} \\
&& B_{1,1}= A_\rho \partial \xi^\rho \qquad B_{1,2}= A_{\rho, \mu} \partial \bX^\rho \xi^\mu \qquad B_{1,3}= {i\over 4} F_{\nu_1,\nu_2, \mu} \xi^{ \mu}\phi^{\nu_1} \phi^{\nu_2} \nonumber \\
&& B_{2,1} ={1\over 2} \lp ( A_{\rho, \mu_1 \mu_2} -A_{\mu_1, \rho \mu_2} ) + i [A_\rho, A_{\mu_1,\mu2}] \rp \partial \bX^\rho \xi^{\mu_1} \xi^{\mu_2} \nonumber  \\
&& B_{2,2}= {i\over 8} \lp F_{\nu_1 \nu_2,\mu_1\mu_2} + i [ F_{\nu_1 \nu_2}, A_{\mu_1, \mu_2}] \rp \phi^{\nu_1} \phi^{\nu_2} \xi^{\mu_1} \xi^{\mu_2} \nonumber \\
&& B_{2,3}= {1\over 2} (A_{\mu_1,\mu_2}- A_{\mu_2,\mu_1}) \partial \xi^{\mu_1} \xi^{\mu_2}\nonumber
\ea
where the first index in each expression above keeps track of the number of bosonic fields $\xi^\mu$ of the vertex. The corresponding vertices appear in Figure \ref{fighetvert}. These vertices appear inside one--loop diagrams shown in Figure \ref{fighet1loop}.

{}For the computation of the diagrams, we use partial integrations in order to move as many derivatives as possible from the bosonic propagators either \\
(a) on the $\psi$-fermionic propagators using the identity
\be\label{derferm}
\partial_{1} K(z_1, z_2)= - \partial_{2} K(z_1, z_2)=- i \pi \apr  \delta^{(2)}(z_1,z_2)
\ee
(b) or on the external $\psi$-fermionic fields using the equations of motion  (\ref{heteqm}).\\
Actually,  diagram (c) is finite while diagram (b) vanishes due to the antisymmetry of the $B_{2,3}$ vertex function. Only the remaining diagrams contribute. For the purpose of making contact with the open string calculation, we use world-sheet representation for the propagators (unlike the momentum space representation of \cite{Sen1985}). We also use similar regularization for the heterotic string propagators as we did for the open string. The result~is
\ba\label{het1loopdiv}
&& \makebox{Figure} \ { \color{blue} 6}=-{1\over 2\pi \apr} G(0) \partial \bX^\rho \psi \bigg ( {1\over 2} \partial^\mu F_{\rho \mu} - {i\over 2} [A_{\rho, \mu}, A_\mu] + i [A_{\rho, \mu}, A_\mu] \\ \nonumber
&& -{i\over 2} [A_{\mu, \rho}, A_\mu] -{1\over 2} (A_\rho A_\mu A^\mu+ A_\mu A^\mu A_\rho) + A_\mu A_\rho A^\mu \bigg )\psi = \\
&& =-{1\over 2\pi \apr} G(0) {1\over 2}\psi  (D^\mu F_{\mu \rho} ) \partial \bX^\rho\psi\ . \nonumber
\ea
This expression gives the same beta function as in the open string case (\ref{open1loopbeta}).

\begin{figure}
\begin{center}
\includegraphics[scale=1]{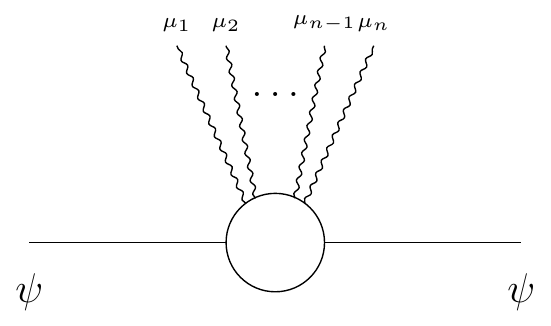}
\caption{Tree-level correlators with two external $\psi$ fermions.\label{figgamma}}
\end{center}
\end{figure}

One could proceed to higher loops  in a similar fashion however, since the interaction vertices are not gauge covariant, the standard procedure is rather cumbersome. We will proceed in a slightly different method where we manage to write perturbation theory in a gauge invariant manner.
In order to compute the part of the beta function involving a single chain of gauge indices (single trace terms), we need correlators with two external $\psi$ fermions, without $\psi$ loops that would contribute additional trace factors.
In order to discuss the loops of $\xi^\mu$ and/or $\phi^\mu$ fields,  it is convenient to start with the tree-level correlators involving two external $\psi$ fields and arbitrary number of $\xi^\mu$ and/or $\phi^\mu$
fields, see Fig.\ref{figgamma}. We will show that, at this level, the correlators can be reorganised in a completely gauge invariant fashion. Consequently, we will obtain a gauge invariant effective action whose diagrammatic rules can be used to compute the loop diagrams required for the beta function.
\begin{figure}
\begin{center}
\includegraphics[scale=0.8]{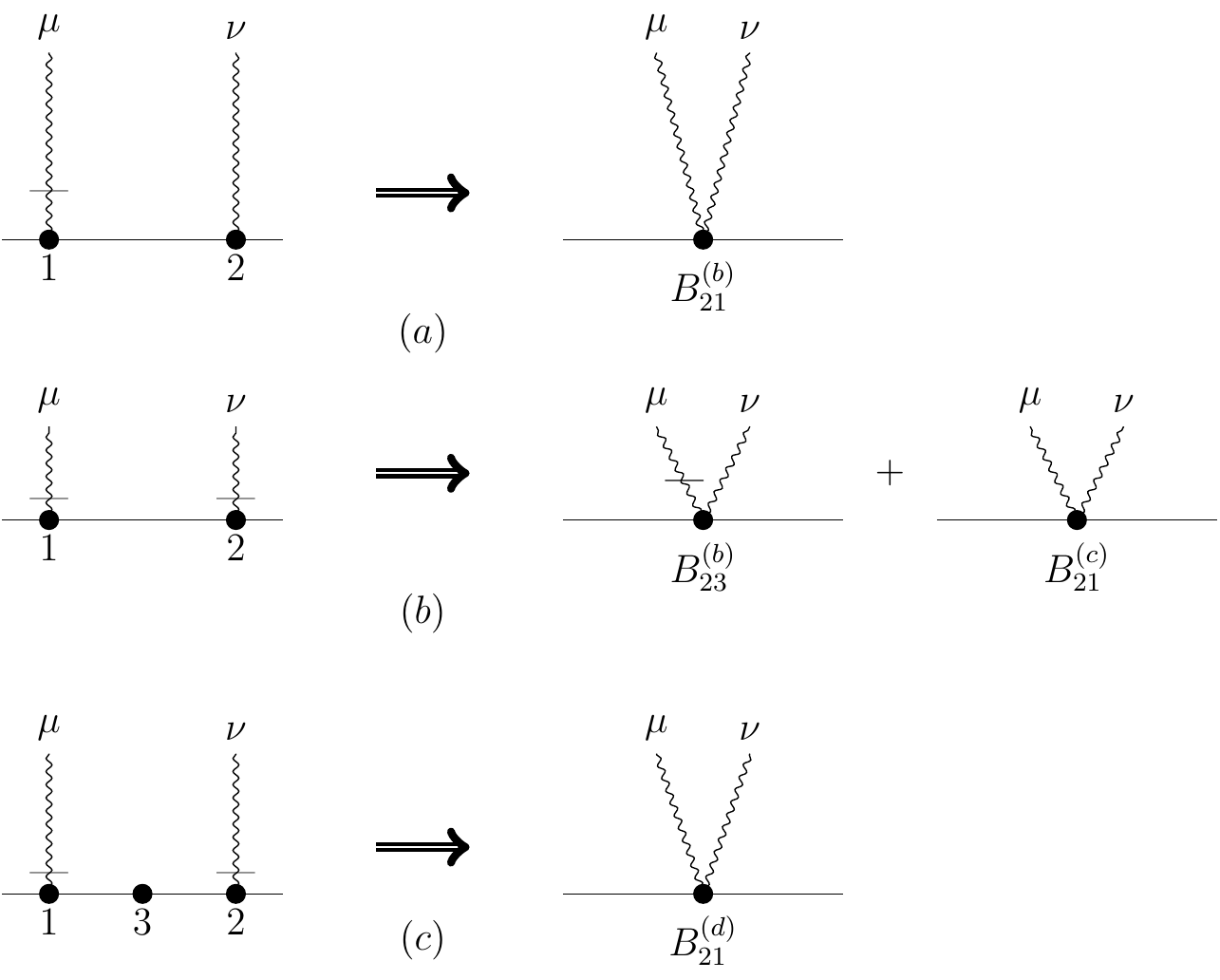}
\caption{\label{figVRxi2}One-particle reducible diagrams combining into effective vertices.}
\end{center}
\end{figure}

First, consider the diagrams which  involve just one $\xi^\mu$ field. It is straightforward to demonstrate that $B_{1,1}$ and $B_{1,2}$ of (\ref{hetvert}) combine, after an integration by parts, in a  gauge invariant form  which can be reproduced by the effective action term
\be\label{hetvertVR1}
+{1\over 2 \pi  \apr} \int d^2 z \ \psi F_{\mu \nu} \xi^\mu \partial \bX^\nu \psi
\ee
Next, consider the correlators with two $\psi$'s and two $\xi^\mu$'s. In addition to the single vertex diagrams from Figure \ref{fighetvert} we have the one-particle reducible  diagrams of Figure \ref{figVRxi2}. After partial integrations, using Eqs. (\ref{derferm}), we obtain
\ba\label{hetvertVR2}
&&B^{(b)}_{21}= -{1\over 2 \apr} i\  [A_{\rho, \mu}, A_\nu] \partial \bX^\rho \xi^\mu \xi^\nu \\
&&B^{(c)}_{21}= -{1\over 2 \apr} {1\over 2}  \lp i [A_{\mu,\rho}, A_\nu] + A_\rho A_\mu A_\nu+ A_\mu A_\nu A_\rho \rp \partial \bX^\rho \xi^\mu \xi^\nu \nonumber \\
&&B^{(d)}_{21}= -{1\over 2 \apr}   A_\mu A_\rho A_\nu \bpsi  \partial \bX^\rho \xi^\mu \xi^\nu  \nonumber \\
&& B^{(b)}_{23}= -{1\over 2 \apr} i \ [A_\mu, A_\nu]   \partial\xi^\mu \xi^\nu \nonumber
\ea
Now combining all of the above along with the vertices from (\ref{hetvert}) and similar expression which include the fermion fields $\phi^\mu$ we obtain the following effective action
\ba\label{VRaction}
&&{\mathcal S}_I= - {1 \over 2 \pi \apr} \int d^2 z\  \bigg( \psi A_\rho \partial \bX^\rho \psi + {i\over 4} \psi ( F_{\mu \nu} \phi^{\mu}\phi^{\nu}) \psi \\
&& + (\psi F_{\mu \nu} \psi) \xi^{\mu} \partial \bX^\nu + {1\over 2} \psi (D_{\mu_1} F_{\mu_2 \rho}) \psi \  \xi^{\mu_1} \xi^{\mu_2} \partial \bX^\rho   \nonumber \\
&& +{i\over 4}\psi (D_{\rho} F_{\mu_1\mu_2})\psi \ \phi^{\mu_1} \phi^{\mu_2} \xi^\rho  +{1\over 2} \psi F_{\mu \nu} \psi \  \xi^{\mu} \partial \xi^\nu \bigg) \nonumber
\ea
The Feynman diagrams constructed by using this action reproduce any correlator with two $\psi$ fields up to order ${\cal O}(\xi^2, \phi^2)$.  Through this section {\it we consider only terms linear in $\partial X^\rho$}. Divergent higher order terms  in $\partial X^\rho$ will be cancelled by exponentiation of diagrams which include counterterms, see analogous discussion in \cite{BP} for the open string.
Modulo the external $\psi$ fields, this action is completely equivalent in structure, albeit integrated on the sphere rather than the disk boundary,  to  ${\cal O}(\xi^2, \phi^2)$ terms of the action in\footnote{We need to make the redefinition
$$ \phi_{het}=  i^{1/2} \sqrt{2} \phi_{open}\ .$$ This is due to the different normalization of the fermion kinetic terms in the open and heterotic action.} (\ref{Wilsonexpansion}). The open string boundary has been substituted by the $\psi$  fields. On the heterotic side the $\psi-\psi$ lines of the Feynman diagrams play the same role as the open string boundary line.

We can construct in a similar fashion the correlators involving two $\psi$ fields and three $\xi^\mu$ fields. This requires vertices from the expansion of the  heterotic action (\ref{hetSI}) to the the third order in the $\xi^\mu$ fields, as well as connected diagrams of lower vertices.
We can then indeed show, by using symmetrized partial integrations\footnote{The most general diagram we can construct to order $\xi^3$ using the gauge choice above, will contain three vertices  $B_{1,1}$ from  (\ref{hetvert}) and there are 3!=6 posssible orderings of  partial integrations. In general to order $\xi^n$ there will be $n!$ orderings of partial integrations} that the generalization of  (\ref{VRaction}) to any order in $\xi$ takes the form
\ba\label{VRactionfinal}
{\mathcal S}_I&=& - {1 \over 2 \pi \apr} \int d^2 z\  [ \psi A_\rho \partial \bX^\rho \psi -{1\over 2} \psi ( F_{\mu \nu} \phi^{\mu}\phi^{\nu}) \psi \\
&& + (\psi F_{\mu \nu} \psi) \xi^{\mu} \partial \bX^\nu + {1\over 2} \psi (D_{\mu_1} F_{\mu_2 \rho}) \psi \  \xi^{\mu_1} \xi^{\mu_2} \partial \bX^\rho   \nonumber \\
&& -{1\over 2}\psi (D_{\rho} F_{\mu_1\mu_2})\psi \ \phi^{\mu_1} \phi^{\mu_2} \xi^\rho  +{1\over 2} \psi F_{\mu \nu} \psi \  \xi^{\mu} \partial \xi^\nu -{1\over 4} \psi D_\rho D_\sigma  F_{\mu\nu} \psi \ \xi^\rho \xi^\sigma \phi^\mu \phi ^\nu   \nonumber \\
&&+\sum_{n=3}^\infty \bigg( {1\over n!}\psi D_{\mu_1} \dots D_{\mu_{n-1}} F_{\mu_n\rho} \psi \ \xi^{\mu_1} \dots  \xi^{\mu_{n}} \partial_\tau \bX^\rho \nonumber \\
&&{n-1 \over n!} \psi D_{\mu_1} \dots D_{\mu_{n-2}} F_{\mu_{n-1}\mu_n}\psi \ \xi^{\mu_1}\dots  \xi^{\mu_{n-1}} \partial_\tau \xi^{\mu_n}  \nonumber \\
&& -{1\over 2} {1\over n!} \psi  D_{\mu_1} \dots D_{\mu_{n}} F_{\rho\sigma}\psi \ \xi^{\mu_1}\dots  \xi^{\mu_{n}} \phi^\rho \phi^\sigma \bigg)] \ .\nonumber
\ea
 By using this action, we can construct, to any loop order, all gauge invariant heterotic diagrams.

  To summarize, we reorganized perturbation theory in a completely background gauge invariant manner so we can proceed to the computation of the two--loop and three--loop beta functions. After this reorganization, the heterotic string perturbation expansion is diagrammatically equivalent to the open string one.

\subsection{Heterotic two--loop beta function}\label{het2loop}
\begin{figure}
\begin{center}
\includegraphics[scale=0.9]{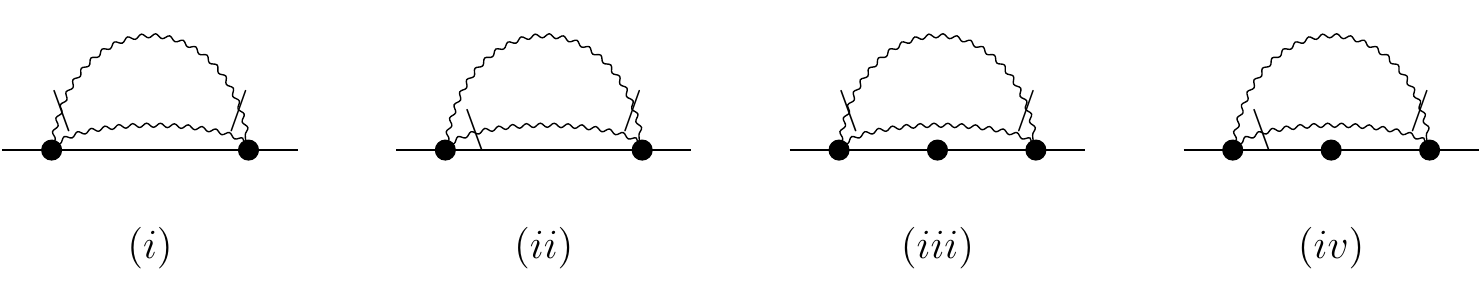}
\caption{\label{fighet2loopextra}Diagrams (a) and (b) of Figure \ref{figopen2loop} now include explicit insertions on $\psi$ lines.}
\end{center}
\end{figure}

At this point, we can use (\ref{VRactionfinal}) to compute the two--loop beta function of the heterotic string. The relevant diagrams are the same as in Figures \ref{figopen2loop}, \ref{figopen2loopb} and \ref{figopen2loopferm}. We start with the bosonic contributions. All the diagrams except (b) and (c) of Figure \ref{figopen2loop} are pretty straightforward. They combine to
\ba \label{het2loopbos}
\ds{-ig  {( \apr)^2 \over 2} \ln^2\epsilon } & \ds{\int}& \ds{  d^2 t \  \psi \lp D^2D^\nu F_{\nu \rho}-2i g[D^\nu F_{\nu \mu},F^{\mu}_{\ \rho}] + {i g \over 2}  [F^{\mu \nu}, D_\rho F_{\mu \nu}] \rp \psi \ \partial_t \bX^\rho } \nonumber \\
\ds{- g^2 { ( \apr)^2 \over 2}  \ln\epsilon} & \ds{\int}& \ds{  d^2 t\  \psi  [F^{\mu \nu}, D_\rho F_{\mu \nu}]\psi \  \partial_t \bX^\rho\ .}
\ea
The first two double pole terms which are proportional to the one--loop beta function (\ref{open1loopbeta}) contribute only to higher order $\apr$ corrections to the beta function and can be ignored.
In the heterotic case, diagrams (b) and (c)  require some extra work. In fact, as shown in  Figure \ref{fighet2loopextra}, there are four diagrams of this type. While diagrams (i) and (ii) are already included as (b) and (c) of Figure \ref{figopen2loop}, diagrams (iii) and (iv) contain $B_{0,1}=A_\rho \partial \widetilde X^\rho$ inserted on the fermion line in the same way as the Wilson loop factor $U[A]$ which is implicitly inserted in (b) and (c).

By using partial integrations, it is easy to show that
\ba \label{het2loopb1}
&& (i)= (ii) +{1\over 2}  G(0)^2\int d^2 z \   ( \psi F_{\mu \nu}) \partial( F^{\mu \nu} \psi)  \nonumber \\
&& (iii)= (iv) -{1 \over 4} G(0)^2 \int d^2 z \   2i \lp \psi F_{\mu \nu} A_\rho \partial X^\rho F^{\mu \nu} \psi  \rp\ .
\ea
After combining these double pole terms and using the field equations (\ref{heteqm}), we find
\ba \label{het2loopb2}
(i)+ (iii)= (ii) + (iv) - {1\over 4} G(0)^2 \int d^2 z \ \psi [D_\rho F_{\mu \nu}, F^{\mu \nu}] \psi\ .
\ea
Next, we compute diagrams (ii) and (iv).
Diagram (ii) gives the following result
\be \label{het2loopiii1}
\int d^2 z_1 \ d^2 z_2 \ \psi(z_1,\bz_1) F_{\mu \nu} (z_1) F^{\mu \nu}(z_2) \psi(z_2,\bz_2) \partial_1 G(z_{12}) \partial_2 G(z_{12}) K(z_{12})\ .
\ee
Define $$ t= {z_1 +z_2 \over 2} \quad u={z_1-z_2\over 2}$$
and expand the fields in $u$
\ba\label{bfexp}
&& A_\mu(z_1,\bz_1)= A_\mu (t) + \partial_\rho A_\mu(t)  \partial_t \bX^\rho(t) \, u + \dots \\[2mm]
&&\psi(z_1 ,\bz_1)= \psi (t , \bar{t}) + \partial_t \psi(t, \bar{t})\, u  + \dots \nonumber
\ea
Finally, change coordinates to polar $u=r e^{i \beta}$, use equations of motion (\ref{heteqm}) and the propagators
\begin{equation}\label{prophet}
G(z_1, z_2)=-{\apr\over 2} \log|z_1-z_2|^2\ , \qquad\quad
K(z_1,z_2)=  - \apr{1\over (\bz_1-\bz_2)}\ .
\end{equation}
After a short calculation we can easily see that the only non-vanishing contribution is
\be \label{het2loopiii2}
 -2(\apr g)^2 \int d^2 t \ (\psi F_{\mu \nu}) \partial( F^{\mu \nu} \psi) \ \int d^2 u\  {1\over |u|^2}\ .
\ee
All other singular terms in the expansion vanish due to symmetry under the angular integration of polar coordinates. The $u$ integral is divergent and gives $2\pi \log \epsilon$ which is the single pole we need for the beta function. Diagram (iv) can be computed in a similar way. In this one we have a triple integral, but the leading and only divergent contribution comes from the region where the three vertices approach each other. We can easily check that this diagram has also a $\log \epsilon$ divergence and gives the terms needed to covariantize the result of diagram (ii). Combining everything we get
\be\label{het2loopb3}
{\color{blue} (b)}=- (\apr g)^2 \lp {1\over 2} \ln \epsilon + {1\over 4} \ln^2\epsilon \rp \int d^2 t \ \psi [ F^{\mu \nu}, D_\rho F_{\mu \nu}] \psi \partial_t \bX^\rho\ .
\ee
As expected, the double pole term cancels the third term of the double poles in (\ref{het2loopbos}).
The single poles of the heterotic bosonic two--loop diagrams combine to give us exactly the same contribution to the beta function  as in (\ref{beta2loopbos}) .

Including the contributions of the $\phi^\mu$ fermion loops  is straightforward. We have the same diagrams as in Figure \ref{figopen2loopferm}. In analogy with the bosonic loops we need to include the equivalent of the Wilson line insertion $U[A]$. The additional diagram looks like (c) of Figure \ref{figopen2loopferm} with $A_\rho \partial \bX^\rho$ inserted between the two vertices. The fermion loop diagrams of the heterotic string, in complete analogy with the open superstring case, cancel the single poles of the bosonic diagrams. As a result, the two--loop  contribution to the beta function vanishes and
we recover the expected heterotic beta function:
\be\label{beta2loophet}
\beta^{het}_\rho={\partial \over \partial(\log \epsilon)}\delta A_\rho= \apr g D^\mu F_{\mu \rho}  + {\cal O}(\apr^3)\ .
\ee
This is consistent with the absence of $F^3$ terms in the heterotic string effective action.

The vanishing of the two--loop constribution to the beta function has been previously demonstrated in Ref. \cite{Fuchs1989} by using superspace techniques. For our purposes, it is preferable though to consider bosonic and fermionic contributions separately and to evaluate world-sheet position integrals instead of momentum integrals. By using this method, it will become easier to make contact with the sv--map.

\section{Single--valued multiple zeta--values and general sv--map proposal for heterotic string}\label{svmapprop}

\subsection{Single--valued multiple zeta--values }

The analytic dependence on the inverse string tension~$\ap$ of string tree--level amplitudes
furnishes an extensive and rich mathematical structure, which is related to modern developments in number theory and arithmetic algebraic geometry, cf.\ Ref. \cite{Stieberger:2016xhs} and references therein.

The topology of the string world--sheet describing tree--level scattering of open strings is
a disk, while tree--level scattering of closed strings is  characterized by a complex sphere.
Open string amplitudes are expressed by integrals along the boundary of the world--sheet disk (real projective line) as
iterated (real) integrals on  $\IR\IP^1\backslash\{0,1,\infty\}$, whose values
(more precisely the coefficients in their power series expansion in $\ap$) are given by multiple zeta values (MZVs)
\be\label{MZV}
\zeta_{n_1,\ldots,n_r}:=\zeta(n_1,\ldots,n_r)=
\sum\limits_{0<k_1<\ldots<k_r}\ \prod\limits_{l=1}^r k_l^{-n_l}\ \ \ ,\ \ \ n_l\in\IN^+\ ,\ n_r\geq2\ ,
\ee
with $r$ specifying the depth and $w=\sum_{l=1}^rn_l$ denoting
the weight of the MZV $\zeta_{n_1,\ldots,n_r}$.
On the other hand, closed string amplitudes  are given by integrals over the complex world--sheet sphere as iterated integrals on
$\IP^1\backslash\{0,1,\infty\}$ integrated independently on all choices of paths.
While in the $\ap$--expansion of open superstring tree--level amplitudes generically the whole space of MZVs \req{MZV} enters  \cite{GRAV,SS}, closed superstring tree--level amplitudes exhibit only a subset of MZVs appearing in their $\ap$--expansion \cite{GRAV,SS}. This subclass can be identified \cite{Stieberger:2013wea} as single--valued multiple zeta values (SVMZVs)
\be\label{trivial}
\SV(n_1,\ldots,n_r)\in\IR
\ee
originating from single--valued multiple polylogarithms (SVMPs) at unity \cite{BrownPoly}.
SVMZVs have  been studied by Brown in \cite{SVMZV} from a mathematical point of view. They have been identified as the coefficients in an infinite series expansion of the Deligne associator
\cite{Deligne} in two non--commutative variables. On the other hand, from a physical point of view SVMZVs
appear in the computation of graphical functions for certain Feynman amplitudes of $\phi^4$ theory \cite{Schnetz}.

The numbers \req{trivial} can be obtained from the MZVs \req{MZV} by introducing the following homomorphism:
\be\label{mapSV}
\sv: \z_{n_1,\ldots,n_r}\mapsto\ \SV( n_1,\ldots,n_r )\ .
\ee
The numbers \req{trivial} satisfy the same double shuffle and associator relations as the
usual MZVs \req{MZV} and many more relations \cite{SVMZV}.
For instance we have (cf. Ref. \cite{Stieberger:2013wea} for more examples):
\begin{align}\label{Example}
\sv(\z_2)&=\SV(2)=0\ ,\\
\sv(\z_{2n+1})&=\SV(2n+1)=2\ \zeta_{2n+1}\ ,\ \ \ n\geq 1\ ,\\
\sv(\z_{3,5})&=-10\ \z_3\ \z_5\ \ \ ,\ \ \ \sv(\z_{3,7})=-28\ \z_3\ \z_7-12\ \z_5^2\ ,\\
\sv(\z_{3,3,5})&=2\ \z_{3,3,5}-5\ \z_3^2\ \z_5+90\ \z_2\ \z_9+\fc{12}{5}\ \z_2^2\ \z_7-\fc{8}{7}\ \z_2^3\ \z_5^2\ ,\ldots\ .
\end{align}

Strictly speaking, the map $\sv$ is defined in the Hopf algebra $\Hc$ of motivic MZVs $\zeta^m$, cf.
\cite{SVMZV} for more details.

In supersymmetric Yang--Mills (SYM) theory a large class of Feynman integrals in four space--time dimensions lives in the subspace of SVMZVs or SVMPs.
As pointed out by Brown in \cite{SVMZV}, this fact opens the interesting possibility
to replace general amplitudes  with their single--valued versions (defined by the map $\sv$),
which should lead to considerable simplifications.
In string theory this simplification occurs by replacing open superstring amplitudes by their
single--valued versions describing closed superstring amplitudes.
In fact, in this work we have detected a large class of Feynman diagrams in two dimensions, which
integrate to SVMZVs by considering heterotic world--sheet beta--functions.

\subsection{General sv--map proposal for heterotic string}

The purpose of the current section is to use the two--loop computation and general results of the preceding sections, in order  to establish a concrete connection between the open string beta function and the heterotic one.
The proposal is the following: assume we can write the beta function of the open string to any loop order as
\be \label{betaopengen}
\beta_o=\sum_n F_n I_n\  |_{\ln \epsilon}
\ee
where the form factors $F_n=F_n(F,D)$  contain the background fields and their covariant derivatives while $I_n$ are ultra-violet divergent integrals of which we keep only the coefficients of single logarithmic divergences needed for the computation of the beta function.  Then we can claim that
\be \label{hetsvopenbeta}
\beta_h= \sum_n F_n H_n \ |_{\ln\epsilon} = \sum_n F_n \ {\rm sv}(I_n)\ |_{\ln \epsilon}={\rm sv}(\beta_o)
\ee
where we used the equivalence of (\ref{VRactionfinal}) with (\ref{Wilsonexpansion}) to show that the factors $F_n$ are the same between the open string and the heterotic string. $H_n$ are the corresponding divergent integrals of the heterotic string. This proposal clearly relies on the existence of a sv--compatible regularization scheme.
Although in the following sections we will concentrate to the lowest derivative terms, $(DF) F^{n-1}$, based on the discussion of the previous section it should be obvious that the proposal is valid for all derivative terms.

Although our discussion is at the level of the sigma model beta functions i.e.\ the equations of motion, the most interesting application is at the level of effective actions. Modulo possible field redefinitions, we expect that the effective action for the single trace gauge sector of the heterotic sting is generated via the sv--map acting on the open superstring effective action.

\section{The sv--map at three loops}\label{het3loop}

{}For the bosonic open string sigma model, the three--loop diagrams of the abelian case have been studied in \cite{DO}.
The world--sheet integrals for the non--abelian diagrams of open string sigma model, unlike the abelian case, are path--ordered. It is  rather difficult to deal with these path-ordered integrals by using traditional momentum space integrals and dimensional regularization.  Instead, we will use world-sheet position integrals and a short-distance cutoff. So when two vertices, at the boundary position $t_1$ and $t_2$, approach each other, there will be a singularity regularized by $|t_1 - t_2| \ge\epsilon$.  In our regularization scheme, the path-ordered integrals can be converted into integrals of hyperlogarithms \cite{HYPERLOG}, which are directly connected to MZVs.

{}For the heterotic string, the three--loop computation appears in \cite{Fuchs1989}, in the framework of superspace and momentum space integrals. Since our goal is to examine the sv--map between  open and heterotic diagrams, we need to use the same regularization scheme for both theories. So for the heterotic diagrams, we will integrate on the complex plane directly, without going to the momentum space. We will use a brute force cutoff: when two vertices at $z_1$ and $z_2$ approach each other, the singularity is regularized by $|z_1 - z_2| \ge \epsilon$.

It is very hard to deal with the full set of diagrams in the nonabelian case, even at three--loops.  So instead of pursuing the complete renormalization program, we have a more modest goal:  for a given Feynman diagram, we want to show that the coefficients of the UV divergent single logarithmic $\ln\epsilon$ terms of the open and heterotic string integrals satisfy the sv--map. This makes sense given the one-to-one correspondence between Feynman diagrams. Nevertheless, such a comparison is quite subtle because a generic diagram contains also  higher powers of logarithms, so the coefficients of single logarithmic terms are regularization dependent, as is any beta function beyond two--loops. The problem essentially boils down to finding a sv--map compatible regularization prescription.

Since we are interested in the single trace terms only, we consider diagrams with one boundary, a single fermion $\psi$ line in the heterotic case, and no $\psi$ loops. We will focus on the diagrams involving bosonic loops. They are non-vanishing at any loop order, although their fermionic counterparts may eventually lead to cancellations.

In the following, we will focus on the diagram shown in Figure \ref{figopen3loopbasic}.
It contributes to the sigma model an ultra-violet divergent Lagrangian term of the form $\partial X^\nu D_{\mu_1} F_{\nu}^{~\mu_3}
F_{\mu_3}^{~\mu_4} F_{\mu_4}^{~\mu_1} $.
\subsection{Open string three--loop integral}
\begin{figure}[b]
\begin{center}
\includegraphics[scale=1.5]{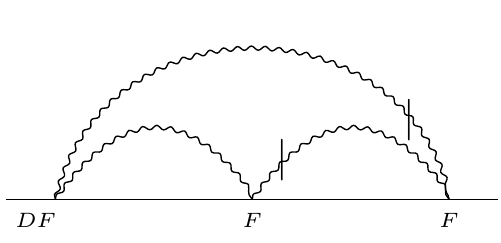}
\caption{The Feynman diagram corresponding to structure $\partial X^\nu D_{\mu_1} F_{\nu}^{~\mu_3}
F_{\mu_3}^{~\mu_4} F_{\mu_4}^{~\mu_1} $.  For the
open string case, the solid line represents the boundary and $F(j)=F(t_j)$. For
the heterotic string case, the solid line represents  the propagator of
$\psi$ and $F(j)=F(z_j)$.\label{figopen3loopbasic}}
\end{center}
\end{figure}
The open string integral associated to the Feynman diagram of Figure \ref{figopen3loopbasic} is given by
\be\label{Ia}
I_{3o}=   \int\limits_{-\infty<t_1<t_2<t_3<\infty}dt_1 dt_2 dt_3\  \frac{\ln{t_{21}}}{t_{31}
    t_{32}}.
\ee
 Short-distance singularities appear when the segments $t_{j+1,j}=t_{j+1}-t_j$ shrink to zero, i.e.\ when the
vertices at $t_j$ and $t_{j+1}$ become coalescent.  With the following change of variables:
\be
    w = t_{31} \ ,\qquad
  u = \frac{t_{21}}{t_{31}} = \frac{t_{21}}{w} \quad (0 < u < 1)\ ,\label{paramo}
\ee
we obtain
\be
  I_{3o} = \int_{-\infty}^{\infty} dt_1   \int_\epsilon^{\mu} \frac{dw}{w} \int_\epsilon^{1-\epsilon} du\
       \frac{\ln{w}  + \ln{u}}{1-u}\label{i3o}\ee
  where we also imposed an infrared cutoff $\mu$ on $w$. We focus, however, on short-distance singularities. The leading $\ln^3\epsilon$ singularity comes from the $\ln w$ term in the numerator.
The single logarithmic term is easy to isolate:
\be  I_{3o}= \int_{-\infty}^{\infty} dt_1 \ \zeta_2\, \ln\epsilon\
        +\dots\label{open2}\ee
       where we neglected higher logarithmic singularities as well as finite terms.
\subsection{Heterotic string three--loop integral}\label{sechet3int}

The heterotic sigma model integral corresponding to the Feynman diagram of Figure \ref{figopen3loopbasic} is
\begin{equation}
  \label{eq:threeloop}
 I_{3h} =  \int \prod_i d^2 z_i \ \prod_{j=1}^{3} \frac{\ln \vert z_{12} \vert^2}{\bar{z}_{12}\bar{z}_{23} z_{23}z_{13}}\ .
\end{equation}
We use the same type of variables as for open strings (\ref{paramo}):
\be
  \label{eq:parametrization}
  z_{31} = r e^{i\theta}\ , \qquad
   \frac{z_{21}}{z_{31}} =x e^{i \alpha}\ .
\ee
 The integral becomes
\be
  \label{eq:integrand}
  I_{3h} = 2\pi\int  d^2z_1 \int \frac{dr}{r} \int dx\,  d\alpha\
    \frac{\ln{r^2} +  \ln{x^2}}{ e^{-i\alpha} |1 - x e^{i\alpha}|^2 }\ . \ee
    Note that $x$ integration covers whole complex plane, $0<x<\infty$, unlike the analogous open string variable $u$ that covers the range $0<u<1$ only. With the complex coordinate system centered at $z_1$, $0<x<1$ corresponds to radial ordering $[z_2,z_3]$ while $1<x<\infty$ to $[z_3,z_2]$.  In order to integrate over the angle $\alpha$, we use the method of Gegenbauer polynomials  outlined in the Appendix, which works not only for one angle, but also when more angles are present at higher loop levels.  This integration  yields different results in the two radial ordering regions. It is convenient to define $u=x^2,~w=r^2$. Then
\be
  \label{eq:angle23}
   I_{3h}= \pi^2 \int  d^2z_1 \int_\epsilon^\mu \frac{dw}{w} \int_0^\infty du  \ln{(wu)} \times\left\{ \begin{array}{lc}\displaystyle
                     \frac{1}{1-u}\ ,& 0<u<1\ ,  \\[3mm]
                   \displaystyle \frac{1}{u(u-1)}\ ,  &u>1\ .
                       \end{array}\right. \ee
The radial ordering $u<1$ yields, up to an overall normalization, the same integral as the open string in  Eq. (\ref{i3o}). It is easy to see that its $\zeta_2\, \ln\epsilon$ part is canceled by the second ordering corresponding to $u>1$. As a result
\be I_{3h}=0 +\dots\ ,\ee
which is consistent with the sv--map, sv$(\zeta_2)=0$.

\section{The sv--map at four loops}\label{het4loops}

We will be considering three representative four--loop  diagrams shown in Figures \ref{fig_open_diagram_1}, \ref{fig_open_diagram_2} and \ref{fig_open_diagram_3}.
The respective contributions to the beta function probe single-trace effective action terms
in which the Lorentz indices of five $F_{\mu\nu}$ tensors are contracted in various ways.
For each diagram, there are four vertices hence three intervals on the open string boundary.
Ultra-violet singularities appear in the limit when the vertices coalesce, i.e.\ when one or more intervals shrink to zero size. With the short distance cutoff $\epsilon$, the logarithmic singularities can be as strong $\ln^4\epsilon$, therefore the single logarithmic terms of interest are very sensitive to the way how this cutoff is imposed, that is to the choice of integration variables and the order in which the intervals shrink to minimum size. Open string positions are real while the heterotic ones are complex, containing radial and angular parts, therefore it is no possible to choose identical variables. There is however, a natural choice of ``matching'' variables, similar to what we used for three--loops, such that after integrating out the angles, one radial ordering of the heterotic string vertex positions, the same as ``time'' ordering on the boundary, yields exactly the same integral as in the open string case. We will show that single value projection appears as a result of adding all radial orderings.\renewcommand{\quad}{~}
\begin{figure}[t]
  \centering
  \includegraphics[width=10cm]{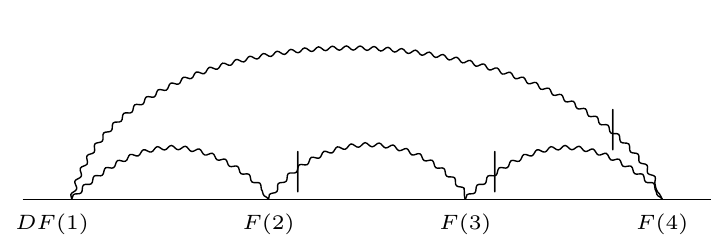}
  \caption{The Feynman diagram contributing to  $\partial X^\nu D_{\mu_1} F_{\nu}^{\quad\mu_3}
F_{\mu_3}^{\quad\mu_4} F_{\mu_4}^{\quad\mu_5} F_{\mu_5}^{\quad\mu_1}$.}
  \label{fig_open_diagram_1}
\end{figure}
\subsection{Figure \ref{fig_open_diagram_1}}\renewcommand{\quad}{~}
 The
vertex structure of this Feynman diagram is $\partial X^\nu D_{\mu_1} F_{\nu}^{\quad\mu_3}
F_{\mu_3}^{\quad\mu_4} F_{\mu_4}^{\quad\mu_5} F_{\mu_5}^{\quad\mu_1}$.
\subsubsection*{(a) Open string}
The open string integral is
\begin{equation}\label{eq:open_diagram_1}
 I_{4o}=\int\limits_{-\infty<t_1<t_2<t_3<t_4<\infty}dt_1 dt_2 dt_3dt_4 \ \frac{\ln{t_{21}}}{t_{41}
    t_{32}t_{43}}\ .
\end{equation}
 With the following change of variables
\be
    w = t_{41} \ ,\qquad~
  u = \frac{t_{21}}{t_{31}}  , \qquad~ v = \frac{t_{31}}{t_{41}} \
  ,\label{param4o}
\ee
we obtain
\be
  I_{4o} = \int_{-\infty}^{\infty} dt_1   \int_\epsilon^{\mu} \frac{dw}{w} \int_\epsilon^{1-\epsilon} dv\int_\epsilon^{1-\epsilon} du\
       \frac{\ln(wu v)}{(1-u)(1-v)}\label{i4o}\ .\ee
It is easy to see that no single logarithmic $\ln\epsilon$ terms apear after integrations. This is consistent with the results of \cite{Medina2002}, where no effective action terms were found corresponding to the respective part of the beta function.
\subsubsection*{(b) Heterotic string}
The heterotic string integral corresponding to this Feynman diagram is
\begin{equation}
  \label{eq:hete_diagram_1}
I_{4h}=\int
 \prod_{j=1}^{j=4} d^2z_j\ \frac{1}{\bar{z}_{12}  \bar{z}_{23} \bar{z}_{34} }\frac{\ln\vert
  z_{12}\vert^2}{z_{23}z_{34} z_{14}}.
\end{equation}
We use the same type of variables as for the open string (\ref{param4o}):
\be
  \label{param4h}
  z_{41} = r e^{i\theta}\ , \qquad
   \frac{z_{21}}{z_{31}} =x e^{i \alpha}\ ,\qquad \frac{z_{31}}{z_{41}} =y e^{i \beta}\ ,
\ee
The integral becomes
\be
  \label{eq:integrand2}
  I_{4h} = 2\pi\int  d^2z_1 \int_\epsilon^\mu \frac{dr}{r} \int dx dy\,  d\alpha d\beta\
    \frac{\ln{r^2} +  \ln{x^2}+  \ln{y^2}}{ e^{-i\alpha}e^{-i\beta} |1 - x e^{i\alpha}|^2  |1 - y e^{i\beta}|^2}\ . \ee
The angular integrals can be performed by using the Gegenbauer method described in the Appendix, by considering each radial ordering region separately. Expressed in terms of $u=x^2,~v=y^2$ and $w=r^2$, the result reads:
\be
  \label{eq:angle41}
   I_{4h}= \pi^3 \int  d^2z_1 \int_\epsilon^\mu \frac{dw}{w} \int du  \int dv \ln (wuv) \times\left\{ \begin{array}{lc}\displaystyle
                     \frac{1}{(1-u)(1-v)}\ ,& ~~u,v\in (0,1)\ ,  \\[3mm]
                   \displaystyle \frac{1}{v(v-1)(1-u)}\ ,  &~~u<1<v\ ,\\[3mm]
                   \displaystyle \frac{1}{u(u-1)(1-v)}\ ,  &~~v<1<u\ ,\\[3mm]
                   \displaystyle \frac{1}{uv(u-1)(v-1)}\ ,  &~~u,v\in (1,\infty)\ .
                       \end{array}\right. \ee
It is easy to see that although all four regions contribute ultraviolet divergent terms, none of them yields a single logarithmic $\ln \epsilon$ term. Hence the diagram under consideration does not contribute to the respective effective action term neither  in open nor in heterotic string theories.
\begin{figure}[b]
  \centering
  \includegraphics[width=10cm]{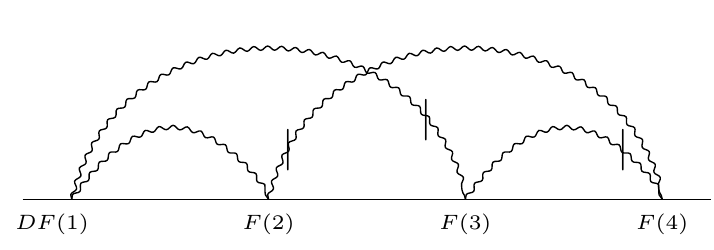}
  \caption{The Feynman diagram contributing to $\partial X^\nu
    D_{\mu_1} F_{\nu}^{\quad\mu_3} F_{\mu_3}^{\quad\mu_4} F_{\mu_5}^{\quad\mu_1}
    F_{\mu_4}^{\quad\mu_5} $ .   }
  \label{fig_open_diagram_2}
\end{figure}\subsection{Figure \ref{fig_open_diagram_2}}
 The
vertex structure of this Feynman diagram is
$\partial X^\nu
    D_{\mu_1} F_{\nu}^{\quad\mu_3} F_{\mu_3}^{\quad\mu_4} F_{\mu_5}^{\quad\mu_1}
    F_{\mu_4}^{\quad\mu_5} $ .
\subsubsection*{(a) Open string}
The open string integral is
\begin{equation}\label{eq:open_diagram2}
 J_{4o}=\int\limits_{-\infty<t_1<t_2<t_3<t_4<\infty}dt_1 dt_2 dt_3dt_4 \ \frac{\ln{t_{21}}}{t_{31}
    t_{42}t_{43}}\ .
\end{equation}
 With the same change of variables as in Eq.(\ref{param4o}), we obtain
\be
  J_{4o}= \int_{-\infty}^{\infty} dt_1   \int_\epsilon^{\mu} \frac{dw}{w} \int_\epsilon^{1-\epsilon} dv\int_\epsilon^{1-\epsilon} du\,
       \frac{\ln(wu v)}{(1-uv)(1-v)}\label{eq:open_3loop_1}\ .\ee
The single logarithmic term is easy to isolate:
\be J_{4o}=
       \int_{-\infty}^{\infty} dt_1 \ \zeta_3\, \ln\epsilon +\dots\label{open3}\ee
       where we neglected higher logarithmic singularities as well as finite terms.
\subsubsection*{(b) Heterotic string}
\begin{figure}[t]
  \centering
  \includegraphics[width=10cm]{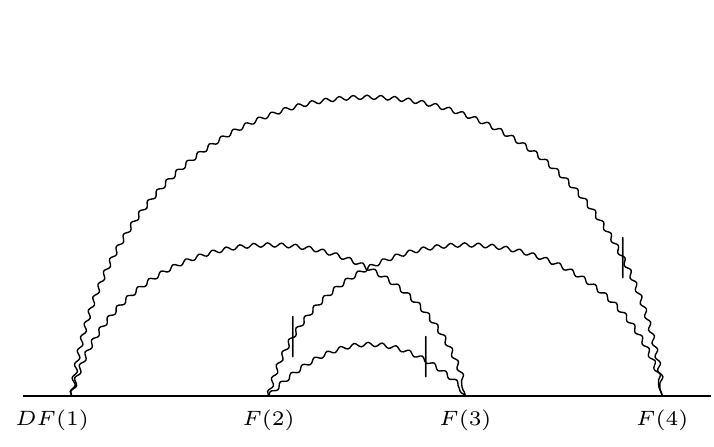}
  \caption{The Feynman diagram corresponding to structure $\partial X^\nu
D_{\mu_1} F_{\nu}^{\quad\mu_3}F_{\mu_4}^{\quad\mu_5} F_{\mu_3}^{\quad\mu_4}
F_{\mu_5}^{\quad\mu_1}$.}
  \label{fig_open_diagram_3}
\end{figure}

The heterotic string integral corresponding to this Feynman diagram is
\begin{equation}
  \label{eq:hete_diagram_2}
J_{4h}=\int
 \prod_{j=1}^{j=4} d^2z_j\ \frac{1}{\bar{z}_{12}  \bar{z}_{23} \bar{z}_{34} }\frac{\ln\vert
  z_{12}\vert^2}{z_{13}z_{24} z_{34}}\ .
\end{equation}
We use the same integration variable as in Eq. (\ref{param4h}), in terms of which
\be
  \label{h13}
  J_{4h} = 2\pi\int  d^2z_1 \int_\epsilon^\mu \frac{dr}{r} \int dx dy\,  d\alpha d\beta\
    \frac{\ln{r^2} +  \ln{x^2}+  \ln{y^2}}{ e^{-i\alpha} e^{-i\beta} (1 - x e^{-i\alpha})  (1 - xye^{i\alpha} e^{i\beta}) |1 - ye^{i\beta}|^2}. \ee
In this case, the angular integrals are slightly harder but can be handled by using the Gegenbauer method described in the Appendix. Depending on six radial orderings, they yield (here again, we use $u=x^2,~v=y^2$ and $w=r^2$):
\begin{eqnarray}
  \label{hf13}
  J_{4h} = \pi^3 \int  d^2z_1&&\!\! \int_\epsilon^\mu \frac{dw}{w} \int du  \int dv\ \ln (wuv)\times\qquad\qquad\qquad\qquad\qquad\qquad\qquad\qquad \\ \nonumber &&
\times  \left\{ \begin{array}{ccl}\displaystyle
                     \frac{1}{(1-u)(1-uv)}\ ,& ~~~~u,v\in (0,1)&\longrightarrow ~\zeta_3\ln\epsilon\ ,  \\[3mm]
                   \displaystyle \frac{1}{v(v-1)(1-u)}\ ,  &~~~~u<1/v<1&\longrightarrow -2\,\zeta_3\ln\epsilon\ ,\\[4mm]
                   \displaystyle 0\ ,  &~~~~v<1/u<1&\longrightarrow ~0\ ,\\[2mm]
                   \displaystyle \frac{1}{uv(1-v)(u-1)}\ ,  &~~~~1/u<v<1&\longrightarrow ~\zeta_3\ln\epsilon\ ,\\[3mm]
                   \displaystyle -\frac{1}{uv(v-1)} \ , &~~~~1/v<u<1&\longrightarrow ~\zeta_3\ln\epsilon\ ,\\[3mm]
                   \displaystyle \frac{1}{uv(v-1)(uv-1)} \ , &~~~~u,v\in (1,\infty)&~\longrightarrow \zeta_3\ln\epsilon\ .
                       \end{array}\right.
  \end{eqnarray}
where for a given ordering, we have also shown the single logarithms emerging after radial integrations. After adding all orderings, we obtain
\be J_{4h}=
       \pi^3\int d^2z_1 \ 2\, \zeta_3 \ln\epsilon +\dots\ .\ee
Comparing with Eq.(\ref{open3}), we find  the result  in agreement with sv$(\zeta_3)=2\zeta_3$.
\subsection{Figure \ref{fig_open_diagram_3}}
\label{sec:diagram_3}
The
vertex structure of this Feynman diagram is
$\partial X^\nu
D_{\mu_1} F_{\nu}^{\quad\mu_3}F_{\mu_4}^{\quad\mu_5} F_{\mu_3}^{\quad\mu_4}
F_{\mu_5}^{\quad\mu_1}$.
\subsubsection*{(a) Open string}
The open string integral is
\begin{equation}\label{eq:open_diagram_3}
K_{4o}= \int\limits_{-\infty<t_1<t_2<t_3<t_4<\infty}dt_1 dt_2 dt_3dt_4\  \frac{\ln{t_{31}}}{t_{41}
    t_{42}t_{32}}\ .
\end{equation}
 With the same change of variables as in Eq.(\ref{param4o}), we obtain
\be
   K_{4o}= \int_{-\infty}^{\infty} dt_1   \int_\epsilon^{\mu} \frac{dw}{w} \int_\epsilon^{1-\epsilon} dv\int_\epsilon^{1-\epsilon} du\
       \frac{\ln(wv)}{(1-uv)(1-u)}\label{eq:open_3loop_2}\ .\ee
It is easy to see that
\be K_{4o} =
       -\int_{-\infty}^{\infty} dt_1 \ \zeta_3\, \ln\epsilon +\dots
       \label{open4}\ee
       where we neglected higher logarithmic singularities as well as finite terms.
\subsubsection*{(b) Heterotic string}
The heterotic string integral corresponding to this Feynman diagram is
\begin{equation}
  \label{eq:hete_diagram_3}
K_{4h}=\int
 \prod_{j=1}^{j=4} d^2z_j \frac{1}{\bar{z}_{12}  \bar{z}_{23} \bar{z}_{34} }\ \frac{\ln\vert
  z_{13}\vert^2}{z_{14}z_{24} z_{23}}\ .
\end{equation}
Using the variables defined in Eq.(\ref{param4h}),
\be  \label{eq:integrand3}
  K_{4h} = 2\pi\int  d^2z_1 \int_\epsilon^\mu \frac{dr}{r} \int dx dy\,  d\alpha d\beta\
    \frac{\ln{r^2} +    \ln{y^2}}{ e^{-i\alpha} e^{-i\beta} (1 - y e^{-i\beta})  (1 - xye^{i\alpha} e^{i\beta}) |1 - xe^{i\alpha}|^2}\ . \ee
After performing angular integrations, depending on six radial orderings, we obtain  (here again, we use $u=x^2, v=y^2$ and $w=r^2$):
\begin{eqnarray}
  \label{eq:angle51}
 K_{4h}  = \pi^3 \int  d^2z_1&&\!\! \int_\epsilon^\mu \frac{dw}{w} \int du  \int dv\ \ln (wv)\times\qquad\qquad\qquad\qquad\qquad\qquad\qquad\qquad \\ \nonumber &&
  \times\left\{ \begin{array}{ccl}\displaystyle
                     \frac{1}{(1-u)(1-uv)}\ ,& ~~~~u,v\in (0,1)&\longrightarrow -\zeta_3\ln\epsilon \ , \\[4mm]
                   \displaystyle  0\ ,&~~~~u<1/v<1&\longrightarrow ~0\ ,\\[2mm]
                   \displaystyle \frac{1}{u(u-1)(1-v)}\ ,   &~~~~v<1/u<1&\longrightarrow -\,\zeta_3\ln\epsilon\ ,\\[3mm]
                   \displaystyle  -\frac{1}{uv(u-1)}\ , &~~~~1/u<v<1&\longrightarrow -\zeta_3\ln\epsilon\ ,\\[3mm]
                   \displaystyle\frac{1}{uv(1-u)(v-1)}\ ,   &~~~~1/v<u<1&\longrightarrow -\zeta_3\ln\epsilon\ ,\\[3mm]
                   \displaystyle \frac{1}{uv(u-1)(uv-1)}\ ,  &~~~~u,v\in (1,\infty)&~\longrightarrow 2\zeta_3\ln\epsilon\ ,
                       \end{array}\right.
  \end{eqnarray}
After adding all orderings, we obtain
\be K_{4h} =
       -\pi^3\int d^2z_1 \ 2\, \zeta_3 \ln\epsilon +\dots\ .\ee
Here again, now comparing with Eq. (\ref{open4}), we find  the result  in agreement with sv$(\zeta_3)=2\zeta_3$.

\section{Conclusions}\label{conclusions}

We addressed the question how the relations between open and heterotic superstring scattering amplitudes discovered in Ref. \cite{Stieberger:2014hba} are reflected by the properties of the effective gauge field theory. According to Ref. \cite{Stieberger:2014hba}, the amplitudes describing the scattering of gauge bosons in both theories, more precisely the amplitudes involving a given single trace gauge group factor, are related by the sv projection which maps open to heterotic amplitudes order by order in the $\alpha'$ expansion. To that end, we studied the
sigma models describing world--sheet dynamics of strings propagating in ambient spacetime endowed with gauge field backgrounds. This framework allows reconstructing the effective action order by order in the string tension parameter $\alpha'$ by studying the beta functions associated to the couplings of background gauge field to the string world--sheet.
The requirement of the vanishing beta functions leads to background field equations generated by the effective action.

{}For open strings, the string coupling to background gauge fields is described by a one dimensional Wilson loop action with the vertex at the boundary. In the heterotic sigma model, the vertices spread over two--dimensional world--sheet. The form of the vertices however, is very similar in both theories. This allows reformulating perturbation theory in terms of identical Feynman diagrams. For each loop correction to the background-string coupling in the heterotic sigma model, there is a matching open string sigma model coupling at the boundary which for a given Feynman diagram is a pull-back of the heterotic one from two to one dimensions. In open string theory, its coefficient is an ordered real integral over vertex positions, while in the heterotic theory the vertices are integrated over whole complex plane. For open strings, the order determines the gauge group factor. In the heterotic case, the gauge group factor is determined by an anti-holomorphic factor hence the vertices come in all radial orderings. This the origin of the difference between the beta functions of these  two sigma models.

We performed explicit three-- and four--loop Feynman diagram computations supporting the conjecture that the beta functions, hence the effective background field actions of open and heterotic superstring theories, are related by the sv map.
The starting point is one-to-one matching between complex and real integrands emerging from a given Feynman diagram.
The position of Feynman vertices on the world-sheet can be parameterized by complex variables that match their real, open string counterparts at the boundary.  We introduced a short-distance cutoff $\epsilon$ and extracted the beta functions from the ultraviolet divergent coefficients of single-logarithmic terms $\sim\ln\epsilon$. In general, these coefficients (hence the beta functions) depend on the details of the cutoff, in particular how the cutoff is imposed on the integration variables. With our choice of variables and cutoffs however,
one of the radial orderings, the canonical one,  yields the same integral as open strings. We showed that the single value projection from open to heterotic string sigma models appears as a result of summing over all radial orderings.

It would be very interesting to find a rigorous proof of a general sv relation between all ordered and complex integrals encountered in Feynman diagrams of sigma models.

\section*{Acknowledgements}
T.R.T. is grateful
to the United States Department of State Bureau of Educational and Cultural Affairs Fulbright
Scholar Program and to Polish-U.S. Fulbright Commission for a Fulbright Award
to Poland.
This material is based in part upon work supported by the
National Science Foundation under Grant Number PHY-1620575.
Any opinions, findings, and conclusions or recommendations
expressed in this material are those of the author and do not necessarily
 reflect the views of the National Science Foundation.

\renewcommand{\thesection}{A}
\setcounter{equation}{0}
\renewcommand{\theequation}{A.\arabic{equation}}
\renewcommand{\thesection}{A}
\setcounter{equation}{0}
\renewcommand{\theequation}{A.\arabic{equation}}
\section*{Appendix: Using Gegenbauer polynomials for angular integrations}\label{math}
Gegenbauer polynomials  $C_m$ appear in the  series expansion \cite{Erdelyi2}:
\be
\fc{1}{(1-2ax+x^2)^p}=\sum_{n=0}^\infty C_n^{(p)}(a)\ x^n\ ,\ |x|<1\ ,\ |a|\leq 1\ .
\ee
{}For our purposes, we need the special case
\begin{align}
  \label{eq:gegenbauerExpand}
  \frac{1}{ \vert 1- x e^{i\alpha} \vert^2} =\frac{1}{  1- 2\cos\alpha+x^2 }= \left\{ \begin{array}{lc}
                     \sum\limits_{n=0}^{\infty} C_n(\cos{\alpha})\ x^n\ , & 0<x<1 \ , \\[4mm]
                   \sum\limits_{n=0}^{\infty}C_n(\cos{\alpha}) \ \displaystyle\frac{1}{x^{n+2}}\ ,  & x>1\ .
                   \end{array}\right.
\end{align}
where (see e.g. \cite{Gradshteyn})
\be\label{App:GegenbauerExp2}
C_n(\cos \alpha)\equiv C_n^{(1)}(\cos \alpha)= \sum_{\scriptstyle k,l\,\ge\, 0\,:\atop \scriptstyle k+l=n}  \cos[ (k-l)\alpha]\ .
\ee
In order to compute angular integrals, we note that
\begin{equation}
  \label{App:GegenbauerIntegral}
  \int_0^{2\pi} d\alpha\  e^{i  p \alpha} C_n(\cos\alpha)=\left\{  \begin{array}{ccc}
                                                                 2\pi\ , & ~~~n=2j+p\, ,& j\ge 0\ , \\
                                                                 0\ , & ~~~~~\rm{otherwise}\ ,&
                                                               \end{array}
                                                             \right.
\end{equation}
which is easy to prove by using elementary methods.

As an example, let us consider the heterotic integral of Figure~\ref{fig_open_diagram_2}, Eq. (\ref{h13}), which contains the following subintegral:
\be
  \label{hsub}
  I_{a} = \int_0^{2\pi} d\alpha \int_0^{2\pi}d\beta\
    \frac{1}{ e^{-i\alpha} e^{-i\beta} (1 - x e^{-i\alpha})  (1 - xye^{i\alpha} e^{i\beta}) |1 - ye^{i\beta}|^2} \ee
First, we expand the integrand in series and use Eq. (\ref{eq:gegenbauerExpand}) to obtain
 \begin{eqnarray}
 I_a &=&\int_0^{2\pi} d\alpha \int_0^{2\pi}d\beta\
 \times\left\{
        \begin{array}{lr}\displaystyle
                          \sum\limits_{l=0}^{\infty} x^l  e^{-l i \alpha}\ ,& ~x<1 \\[4mm]\displaystyle
                          \sum\limits_{l=0}^{\infty} \displaystyle - \frac{1}{x^{l+1} } e^{(l+1) i \alpha}\ ,& ~x>1
                        \end{array}
                        \right\} \times\nonumber  \\
                        {} & \times&
 \left\{
          \begin{array}{lr}\displaystyle
                   \sum\limits_{m=0}^{\infty} x^m y^m e^{m i \alpha} e^{m i\beta}\ ,& ~xy < 1 \\[4mm] \displaystyle
                  \sum\limits_{m=0}^{\infty} - \frac{1}{x^{m+1} y^{m+1}} e^{-(m+1) i \alpha} e^{- (m+1)i\beta}\ ,& ~xy >1
                        \end{array}
                        \right\}\times \\ \nonumber && ~~~~~~~~~~~~~~~~\times
 \left\{
           \begin{array}{lr} \displaystyle
                          \sum\limits_{n=0}^{\infty} C_n(\cos{\beta}) y^n\ ,& ~y<1\\[4mm] \displaystyle
                          \sum\limits_{n=0}^{\infty}  C_n(\cos{\beta}) \frac{1}{y^{n+2}}\ ,& ~y>1
                        \end{array}
                         \right\}\ .
  \end{eqnarray}
Next, we perform angular integrals by using  Eq. (\ref{App:GegenbauerIntegral}). Finally, after resumming the resulting series, we obtain Eq. (\ref{hf13}).

\end{document}